\documentclass[10pt]{article}
\usepackage{ametsoc2col}
\newcommand{\myabstract}{Equations for the wave-averaged three-dimensional momentum equations have been 
published in this journal. It appears that these equations are not consistent with the known depth-integrated 
momentum balance, especially over a sloping bottom. These equations should thus be considered with caution 
as they can produce erroneous flows, in particular outside of the surf zone. It is suggested that the inconsistency in the
equations may arise from the different averaging operators applied to the different terms of the momentum equation. It is concluded 
that other forms of the momentum equations, expressed in terms of the quasi-Eulerian velocity, are better suited for 
three dimensional modelling of wave-current interactions.
}
\begin{document}
%
%%%%%%%%%%%%%%%%%%%%%%%%%%%%%%%%%%%%%%%%%%%%%%%%%%%%%%%%%%%%%%%%%%%%%
% TITLE
%
% Enter your TITLE here
%%%%%%%%%%%%%%%%%%%%%%%%%%%%%%%%%%%%%%%%%%%%%%%%%%%%%%%%%%%%%%%%%%%%%
\title{\textbf{\large{Comments on ``The Depth-Dependent Current and Wave Interaction Equations: A Revision''}}}
%
% Author names, with corresponding author information. 
% [Update and move the \thanks{...} block as appropriate.]
%
\author{\textsc{Anne-Claire Bennis,}\\
\centerline{\textit{\footnotesize{Universit\'e Bordeaux 1, CNRS, UMR 5805-EPOC, F-33405 Talence, France}}}\\
\centerline{\textsc{Fabrice Ardhuin}\thanks{\textit{Corresponding author address:} 
				Fabrice Ardhuin, 
				Ifremer, Centre de Brest, 29200 Plouzan{\'e}. 
				\newline{E-mail: ardhuin@shom.fr}}
}\\% Add additional authors, different insitution
\textit{\footnotesize{Ifremer, Laboratoire d'Oc{\'e}anographie Spatiale, Centre de Brest, 29200 Plouzan{\'e}, France}}
%\and 
%\centerline{\textsc{Extra Author}}\\% Add additional authors, different insitution
%\centerline{\textit{\footnotesize{Affiliation, City, State/Province, Country}}}
}
%
% Formatting done here...Authors should skip over this.  See above for abstract.
\ifthenelse{\boolean{dc}}
{
\twocolumn[
\begin{@twocolumnfalse}
\amstitle

% Start Abstract (Enter your Abstract above.  Do not enter any text here)
\begin{center}
\begin{minipage}{13.0cm}
\begin{abstract}
	\myabstract
	\newline
	\begin{center}
		\rule{38mm}{0.2mm}
	\end{center}
\end{abstract}
\end{minipage}
\end{center}
\end{@twocolumnfalse}
]
}
{
\amstitle
\begin{abstract}
\myabstract
\end{abstract}
}
%%%%%%%%%%%%%%%%%%%%%%%%%%%%%%%%%%%%%%%%%%%%%%%%%%%%%%%%%%%%%%%%%%%%%
% MAIN BODY OF PAPER
%%%%%%%%%%%%%%%%%%%%%%%%%%%%%%%%%%%%%%%%%%%%%%%%%%%%%%%%%%%%%%%%%%%%%
\section{Introduction}
The wave-averaged conservation of momentum can take essentially two forms, one for the mean flow momentum only, and the 
alternative form for the full momentum, which includes  the wave pseudo-momentum  \citep[hereinafter `wave momentum', see][]{McIntyre1981}. This question is well known for depth-integrated equations \citep{Longuet-Higgins&Stewart1964,Garrett1976,Smith2006b}, but the vertical profiles of 
the mass and momentum balances are more complex. The pioneering effort of \citet[][hereinafter M03]{Mellor2003} produced practical 
wave-averaged for the total momentum that, in principle, may be used in primitive equation models to investigate coastal flows, such as the 
wave-driven circulations observed by \cite{Lentz&al.2008}. The first formulation \citep{Mellor2003} was slightly inconsistent due to the 
improper approximation of wave motion with Airy wave theory, which is not enough on a sloping bottom, however small the slope may be. 
This question was discussed by \cite{Ardhuin&al.2008}, and a correction was given and verified. These authors acknowledged that 
these equations, when using the proper approximation, are not well suited for practical applications because very complex wave models 
are required for the correct estimation of the vertical fluxes of wave momentum, that are part of the fluxes of total momentum. 

Although M03 gave correct wave-forcing expressions -- in terms of velocity, pressure and wave-induced displacement, before any approximation -- 
\citet[][hereinafter M08]{Mellor2008b} derived a new and different solution from scratch.
The two theories may be consistent over a flat bottom, but they differ at their lowest order over sloping bottoms, so that the M08 equations are likely to be 
flawed, given the analysis of M03 by \cite{Ardhuin&al.2008}, and the fact that their consistency  was not verified numerically over sloping bottoms. 

Instead, M08 asserted that the equations are consistent with the depth-integrated equations of \cite{Phillips1977}. Further, 
about the test case proposed by \cite{Ardhuin&al.2008}, M08 stated that the 
wave energy was unchanged along the wave propagation and that the resulting wave forcing whould be uniform over the depth. 
Here we show that the M08 equations do not yield the known
depth-integrated equations \citep{Phillips1977}
with a difference that produces very different mean sea level variations when waves propagate over a sloping bottom. As for the 
test case of proposed by \cite{Ardhuin&al.2008}, we show that a consistent analysis should take into account the small but 
significant change in wave energy due to shoaling. In the absence of dissipative processes,
 the M08 equations can produce spurious velocities of at least 30~cm/s, with 1~m high waves over a bottom slope of the order of 1\% in 4~m water depth. 

\section{Depth-integration of the M08 equations}
 For simplicity we consider motions limited to a vertical plane $(x,z)$ with constant water density 
and no Coriolis force nor wind stress or bottom friction. The wave-averaged momentum equation in M08
takes the form
\begin{equation}
\frac{\partial U}{\partial t}+\frac{\partial U^2}{\partial x}+ \frac{\partial U W}{\partial z} = 
-g \frac{\partial \widehat{\eta}}{\partial x}  + F,\label{M08_U}
\end{equation}
and the continuity equation is
\begin{equation}\label{M08_CONT}
\displaystyle\frac{\partial U}{\partial x}+\frac{\partial W}{\partial z}=0.
\end{equation}

Where $U$ and $W$ are the Lagrangian mean velocity components, which contains the current and Stokes drift velocities, $g$ is the 
acceleration due to gravity and $ \widehat{\eta}$ is the time-averaged water level at the horizontal position $x$. 
The force given by M08 on the right hand side of (\ref{M08_U}) can be written as the sum 
\begin{equation}
F=F_{px}^{M08}+F_{uu}
\end{equation}
of a wave-induced pressure gradient
\begin{eqnarray}
F_{px}^{M08}&= & - \frac{\partial S_{px}^{M08}}{\partial x }  =  - \frac{\partial }{\partial x } \left(E_D -\overline{\widetilde{w}^2}\right) \\ 
& \simeq &  - \frac{\partial}{\partial x } \left( E_D - kE F_{SC}F_{SS}\right)
\end{eqnarray}
 and  the divergences of the horizontal flux of wave momentum,
\begin{eqnarray}
F_{uu}^{M08}&= &- \frac{\partial S_{uu}}{\partial x } = - \frac{\partial \overline{\widetilde{u}^2}}{\partial x} \\ 
& \simeq &  - \frac{\partial }{\partial x}\left(kE F_{CC}F_{CS} \right)
\end{eqnarray}
where $E$ is the wave energy, $k$ is the wavenumber, $\widetilde{u}$ and $\widetilde{w}$ are respectively the horizontal and vertical wave-induced (orbital) 
velocities. 
$E_{D}$ is defined by
\begin{equation}
E_{D}=0 \hskip5pt \hbox{if} \hskip5pt z\neq\widehat{\eta}\hskip5pt \hbox{and} \int_{-h}^{\widehat{\eta}^{+}} E_{D}dz=\frac{E}{2}. 
\end{equation}

Using the  mean water depth $D$, and bottom elevation $-h$, $F_{CC}$, $F_{SS}$ and $F_{SC}$ are non-dimensional functions of $kz$ and $kD$, 
\begin{eqnarray}
F_{CC}&=&\frac{\cosh(kz+kh)}{\cosh(kD)}, \\
F_{SS}&=&\frac{\sinh(kz+kh)}{\sinh(kD)}, \\
F_{SC}&=&\frac{\sinh(kz+kh)}{\cosh(kD)}. 
\end{eqnarray}

%%%%%%%%%%%%%%%%%%%%%%%%%%%%%%%%%%%%%%%%%%%%%%%%%%%%%%%%%%%%%%%%%%%%%
\begin{figure*}[htb]
     \noindent\centerline{\includegraphics[width=0.8\linewidth,angle=0]{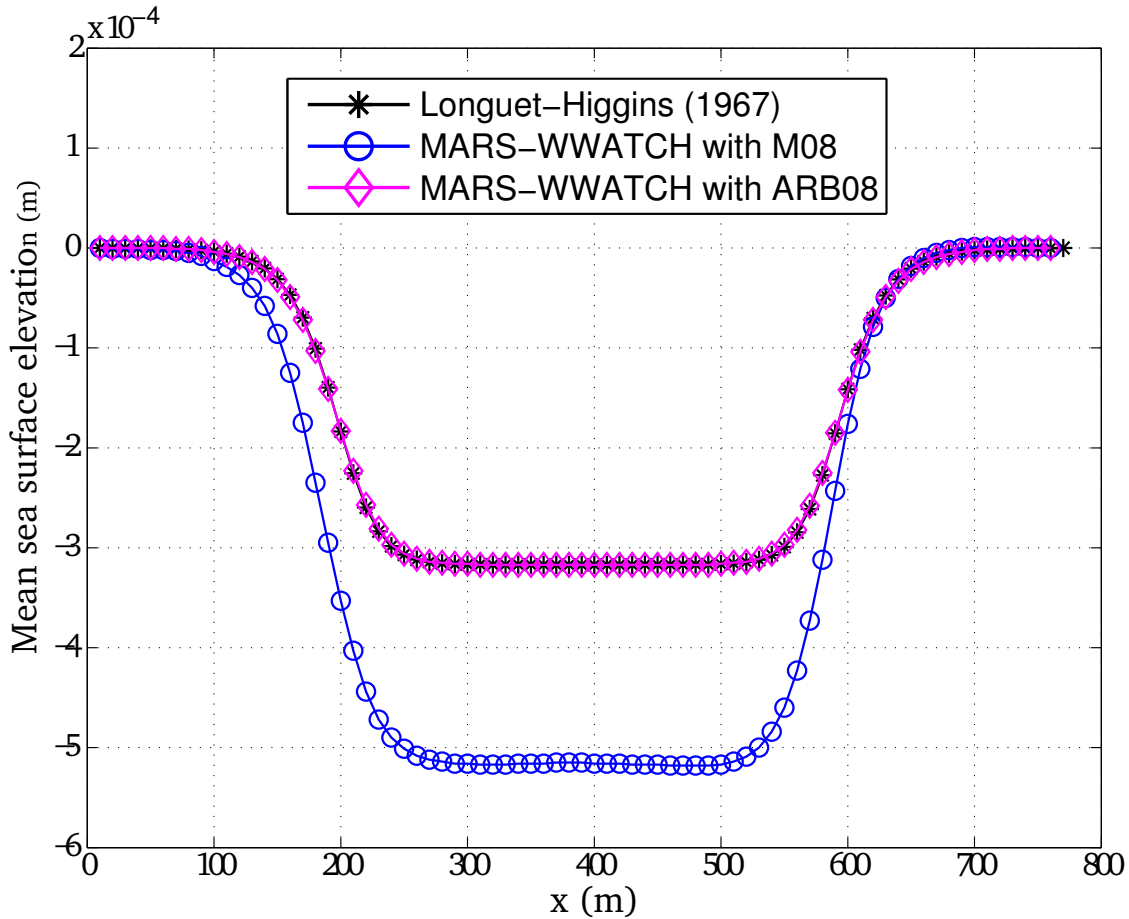}}\\
 %  \noindent\includegraphics[width=\linewidth,angle=0]{./figures/comp-xe-LH-M08-ARB08.pdf}\\
  \caption{Mean sea surface elevation induced by monochromatic waves propagating over the smooth bottom shown in figure \ref{f2}, with amplitude
 $H_{s}=0.34$~m and period $T=5.24$~s. The extra terms forcing terms in eq. (\ref{extra}) lead to an overestimation of the set-down by more than 50 \%  for this case.
ARB08 stands for quasi-Eulerian momentum equations of \cite{Ardhuin&al.2008}.}\label{f1}
\end{figure*}
%%%%%%%%%%%%%%%%%%%%%%%%%%%%%%%%%%%%%%%%%%%%%%%%%%%%%%%%%%%%%%%%%%%%%
The depth-averaged mass-transport velocity is \begin{equation}
\displaystyle\overline{U}=\frac{1}{D}\int_{-h}^{\widehat{\eta}} U {\mathrm d} z.                                              
                                              \end{equation}
M08 correctly noted that 
\begin{equation}
\int_{-h}^{\widehat{\eta}} \underbrace{\left(S_{px}^{M08} + S_{uu}\right)}_{\displaystyle=S_{xx}^{M08}} {\mathrm d} z   =S_{xx}^{P77}                     
                         \end{equation}
with $S_{xx}^{P77}$ given by 
\cite{Phillips1977}. However, for a depth-uniform $U$, the depth integrated momentum equation in \cite{Phillips1977} is
\begin{equation}
 \frac{\partial (D\overline{U})}{\partial t}+ \frac{\partial }{\partial x}\left(D\overline{U}^2\right) = 
-g D \frac{\partial \widehat{\eta}}{\partial x}  + \frac{\partial }{\partial x} S_{xx}^{P77}.     \label{P77}
\end{equation}

The forcing in the depth-integration of (\ref{M08_U}) differs from the forcing in (\ref{P77}), because the gradient is inside of the integral, namely, 
\begin{eqnarray}
\int_{-h}^{\widehat{\eta}} \frac{\partial S_{xx}^{M08}}{\partial x} dz&=&\frac{\partial S_{xx}^{P77}}{\partial x} - S_{xx}^{M08}(z=-h) \frac{\partial h}{ \partial x}  \nonumber \\
 & & -  S_{xx}^{M08}(z=\widehat{\eta}) \frac{\partial \widehat{\eta}}{ \partial x}.\label{extra}
\end{eqnarray}
The depth integral of M08 thus includes two extra term. In particular $S_{xx}^{M08}(z=-h) \partial h / \partial x=-2 kE  (\partial h / \partial x) /  \sinh(2kD)$ 
can be dominant over a sloping bottom. 
As a result the momentum balance in M08, unlike M03, does not produce the known set-down and set-up. This is illustrated in figure \ref{f1}. 
We take the case proposed by \cite{Ardhuin&al.2008} with steady monochromatic 
waves shoaling on a slope without breaking nor bottom friction and for an inviscid fluid, conditions in which exact numerical solutions are known.  
The bottom slopes 
smoothly from a depth $D=6$ to $D=4$~m. We added a symmetric slope back down to 6~m to allow periodic boundary conditions if needed.
For a wave period of 5.24~s the group velocity varies little from $4.89$ to 4.64~m~s$^{-1}$, giving a 2.7\% increase of wave amplitude on the shoal.
Contrary to statements in M08, $\partial E/\partial x$ is significant, with a 5.4\% change of $E$ over a few wavelengths.  

From the Eulerian analysis of that situation \citep[e.g.][]{Longuet-Higgins1967}, the mean water level 
should be 0.32~mm lower on the shoal (figure \ref{f1}). \cite{Rivero&Arcilla1995}  
established that there is no other dynamical effect: a steady Eulerian mean current develops, compensating for the 
divergence of the wave-induced mass transport \citep[see also][]{Ardhuin&al.2008}. 

\section{Flows produced by the M08 equations}
\begin{figure*}[htb]
    \noindent\centerline{\includegraphics[width=0.8\linewidth,angle=0]{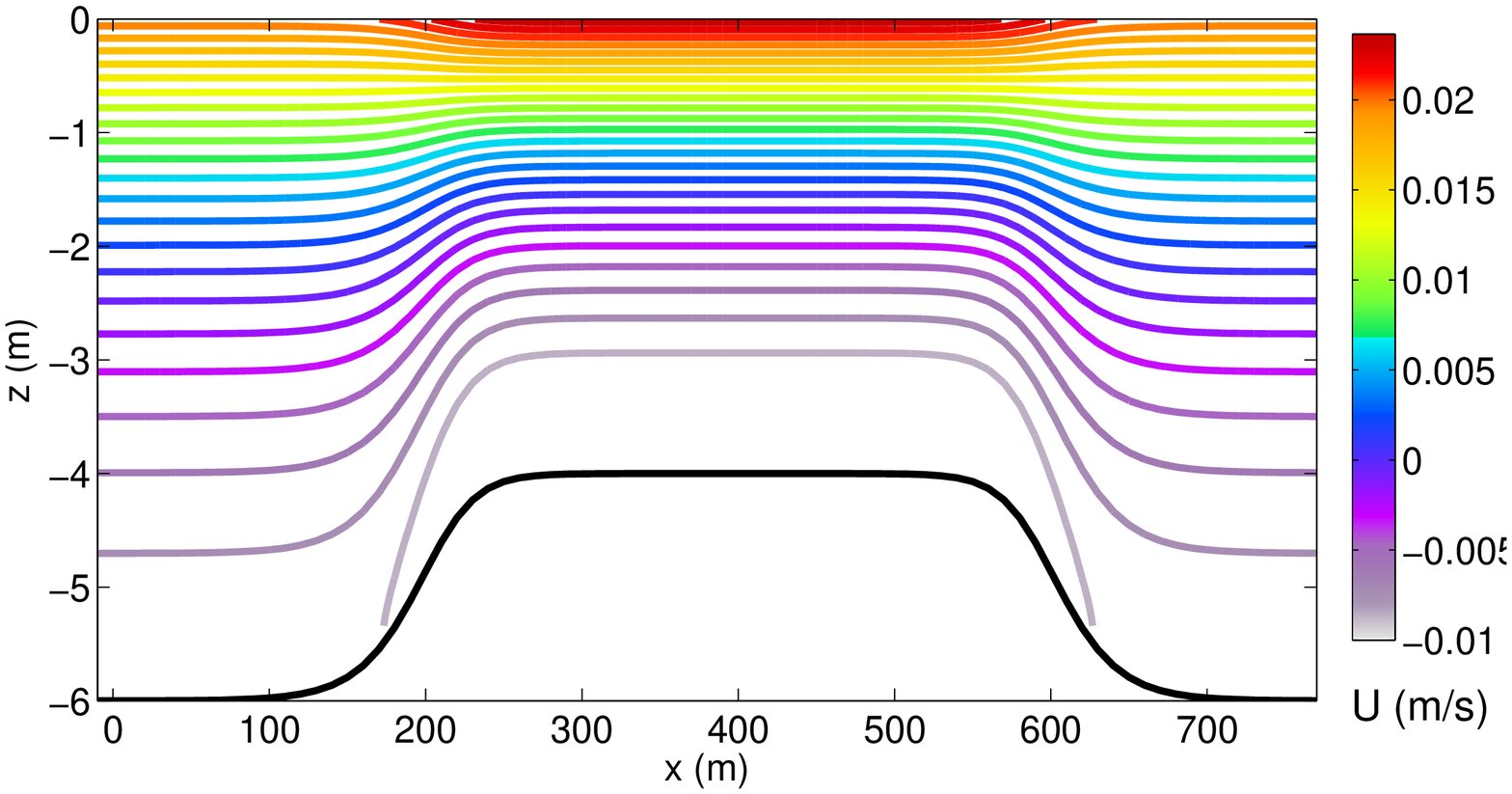}}\\
%    \noindent\includegraphics[width=\linewidth,angle=0]{./figures/UL_matlab.pdf}\\
  \caption{Lagrangian velocity $U$ for the inviscid sloping bottom case with $H_{s}=1.02$~m and $T=5.24$~s, obtained from the 
quasi-Eulerian analysis as $U=\widehat{u}+U_s$. Contours are equally spaced from -0.01 to 0.025 m~s$^{-1}$. The thick black 
line is the bottom elevation.}\label{f2}
\end{figure*}

Because the relative variation in phase speed is important, 
from 6.54 to 5.65 ~m~s$^{-1}$, the Stokes drift accelerates on the shoal. 
The Eulerian velocity $\widehat{u}$ is irrotational, thus nearly depth-uniform, and compensates the Stokes drift divergence
 by a convergence. The Lagrangian velocity $U$, shown in figure \ref{f2}, is the sum of the two steady 
velocity fields. 
%%%%%%%%%%%%%%%%%%%%%%%%%%%%%%%%%%%%%%%%%%%%%%%%%%%%%%%%%%%%%%%%%%%%%
%%%%%%%%%%%%%%%%%%%%%%%%%%%%%%%%%%%%%%%%%%%%%%%%%%%%%%%%%%%%%%%%%%%%%

We now solve for the equations derived by M08. 
The numerical solution is obtained by coupling the WAVEWATCH~III wave model \citep{Tolman2009}, 
solving the phase-averaged wave action equation, 
and the MARS3D ocean circulation model \citep{Lazure&Dumas2008}. This coupling
uses the generic coupler PALM \citep{Buis&al.2006}. The feedback 
from flow to waves is negligible here and was thus turned off. MARS3D was implemented with 100 sigma levels regularly spaced, 
and 5 active points in the transversal $y$ direction,
 with 2 extra wall points, 
and 2 ghost points needed to define finite differences, it is thus a reall three-dimensional calculation although the physical situation is 
two-dimensional. There are 78 active points in the $x$ direction. The time step was set to 
0.05~s for tests with $H_s=1.02$~m  (1~s for $H_s=0.34$~m). For simplicity, the wave model forcing is updated at each time step. 
We use Eq. (\ref{M08_U}) transformed to $\varsigma$ coordinates, with $\varsigma$ defined by
 $z=s(x,\varsigma,t)=\widehat{\eta}+\varsigma D + \widetilde{s}$ \citep{Mellor2003}, 
\begin{equation}
\frac{\partial U}{\partial t}+U\frac{\partial U}{\partial x}+\frac{W}{D} \frac{\partial U}{\partial \varsigma} = F-g\frac{\partial \widehat{\eta}}{\partial x}.
\label{Melloreq}
\end{equation}
where the advection terms are obtained by using Eq. (\ref{M08_CONT}).

The flow boundary conditions are open. 
The monochromatic wave amplitude $a=0.12$~m translates into a significant wave height $H_s$ of $0.34$~m for random 
waves with the same energy.  We also test the model with  $a=0.36$~m, i.e. 
$H_s=1.02$~m, still far from the breaking limit in 4~m depth. 
%%%%%%%%%%%%%%%%%%%%%%%%%%%%%%%%%%%%%%%%%%%%%%%%%%%%%%%%%%%%%%%%%%%%%
\begin{figure}[t]
     \noindent\includegraphics[width=\columnwidth]{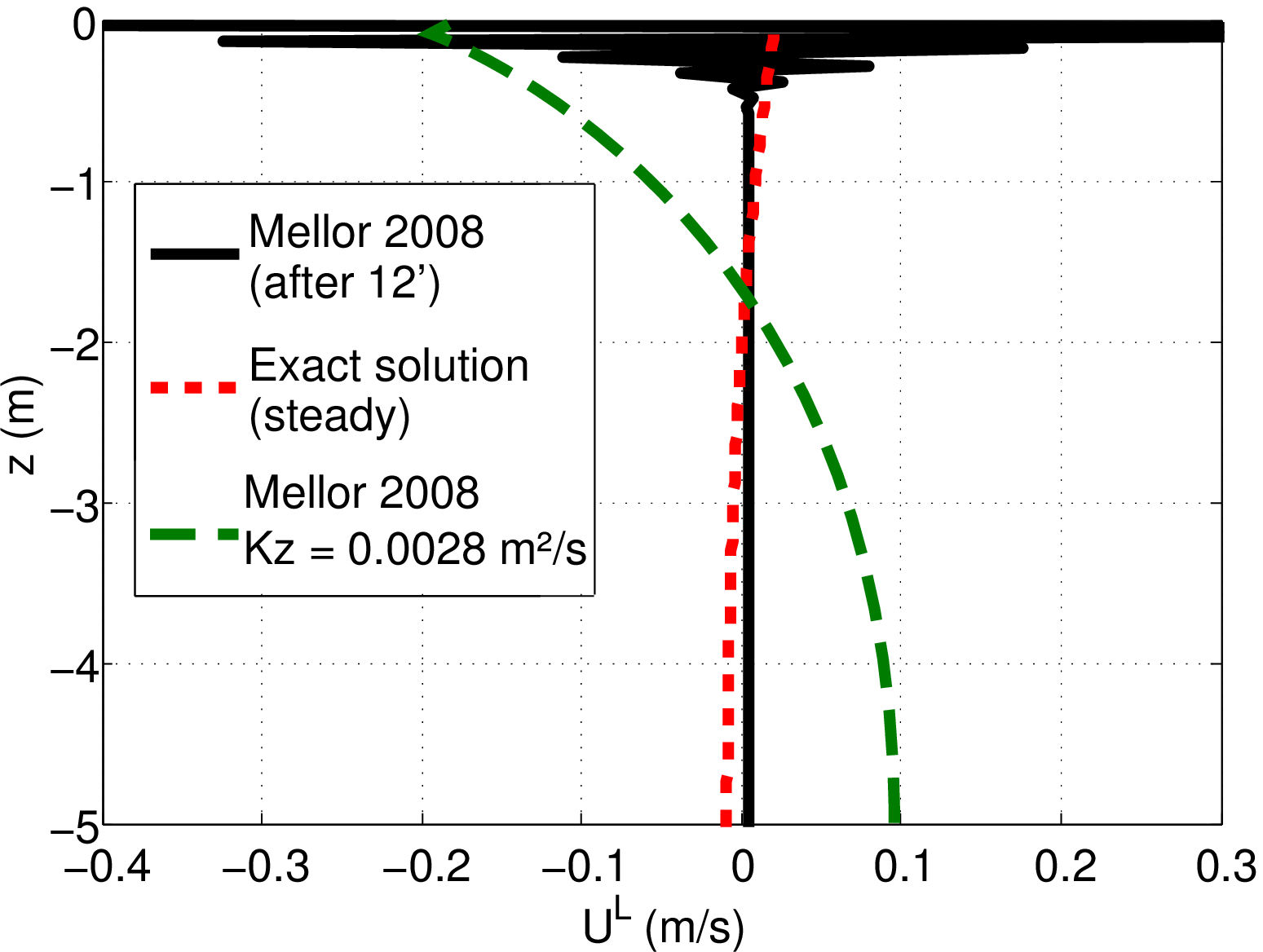}
%   \noindent\includegraphics[width=\columnwidth]{./figures/M08-profiles-inv-FA.pdf}
\caption{\label{CVP}Comparison of vertical profiles of $U$ at $x=200$~m given by different models: M08 without mixing (solid black line), 
M08 with mixing (dashed green line), exact solution (dashed red line). 
The wave parameters are $H_{s}=1.02$~m and $T=5.26$~s. All profiles are plotted after six minutes of time integration. The $x$-axis was clipped, 
and the maximum velocities with M08 reached 0.8 m~s$^{-1}$.}
\end{figure}
%%%%%%%%%%%%%%%%%%%%%%%%%%%%%%%%%%%%%%%%%%%%%%%%%%%%%%%%%%%%%%%%%%%%%

The discontinuity  of the vertical profile in the forcing $F$, due to the $E_D$ term, is not easily ingested 
by the numerical model, and generates a strongly oscillating 
velocity profile (Fig. \ref{CVP}). These oscillations are absent at depths larger than 0.8~m, consistent 
with the zero values of $F$ below the surface. A realistic constant viscosity $K_z=2.8.10^{-3}$~m$^{2}$~s$^{-1}$ removes 
the oscillations by diffusing the negative  term $-\partial E_D/\partial x$ over the vertical. Yet this term is a momentum source that produces 
velocities one order of magnitude larger than 
the Stokes drift $U_s$, with an opposite sign (Fig. \ref{CVP}). The spurious velocities given by M08 with a realistic mixing 
are  most pronounced for waves in not too shallow water (Table \ref{t2}), and comparable with those given by the M03 equations without mixing. 
%%%%%%%%%%%%%%%%%%%%%%%%%%%%%%%%%%%%%%%%%%%%%%%%%%%%%%%%%%%%%%%%%%%%%%%%%%%%%%%%%%%%%%%%%%%%%%%%%%%%%%%%%%%%%%%%%%%%%%%%%%%%%%%%%
\begin{table}[t]\label{TM08}
\caption{\textbf{Model results with Mellor (2008):} Surface velocity at $x=$200~m (on hte up-slope) for different model settings. The settings corresponding to the test
 in \cite{Ardhuin&al.2008} are given in the second line. The surface
 velocity values are written for the time $t=900$~s except for the case without mixing ($t=360$~s).  }\label{t2}
\begin{center}
\begin{tabular}{ccccc}
\hline\hline
$H_s$(m) &$T_p$(s) & $K_z$~(m$^{2}$~s$^{-1})$ & $U$~(m~s$^{-1})$ \\
\hline
 1.02 & 5.6 & 0 & 0.6116 \\
 0.34 & 5.6 & 0 & 0.2127  \\
 0.34 & 13 & 0 & 0.3164  \\
 1.02 & 5.6 & 2.8$.10^{-3}$ & -0.1594  \\
 0.34 & 5.6 & 2.8$.10^{-3}$ & -0.0256  \\
 0.34 & 13 & 2.8$.10^{-3}$ & -0.0007 \\
\hline
\end{tabular}
\end{center}
\end{table}

\section{Conclusions}
We showed that the equations derived by \cite{Mellor2008b} appear inconsistent with the know depth-integrated momentum balances in the presence of a sloping bottom. 
In the absence of dissipation, a numerical integration of these equation produce unrealistic surface elevations and currents. The currents may reach significant values for
very moderate waves, exceeding the expected results by one order of magnitude. 
While we did not discuss the origin of the inconsistency, it appears that \cite{Mellor2008b} used a different averaging for the pressure gradient term 
and for the advection terms of the same equation. We believe that this is the original reason for the problems discussed here. 
The spurious velocities produced by M08 are likely to be dwarfed by the strong forcing 
imposed by breaking waves in the 
surf zone. Nevertheless, we expect that the M08 equations can produce large errors for continental shelf applications, such as the investigation of cross-shore transports outside of the surf zone. 
Alternatively, equations for the quasi-Eulerian velocity can be used \cite{McWilliams&al.2004,Ardhuin&al.2008,Uchiyama&al.2009} which do not have such problems.

\begin{acknowledgment} 
A-C. B. acknowledges the support of a post-doctoral grant from INSU and grant ANR-BLAN-08-0330-01. F.A. 
is supported by a  FP7-ERC grant \#240009 ``IOWAGA'', and the U.S. National Ocean Partnership Program, under ONR grant N00014-10-1-0383.
\end{acknowledgment}

% Use appendix}[A], {appendix}[B], etc. etc. in place of appendix if you have multiple appendixes.
\ifthenelse{\boolean{dc}}
{}
{}
%{\clearpage}

% Create a bibliography directory and place your .bib file there.
\ifthenelse{\boolean{dc}}
{}
{\clearpage}
\bibliographystyle{./ametsoc}
\bibliography{../../references/wave}
%%%%%%%%%%%%%%%%%%%%%%%%%%%%%%%%%%%%%%%%%%%%%%%%%%%%%%%%%%%%%%%%%%%%%
% FIGURES
%%%%%%%%%%%%%%%%%%%%%%%%%%%%%%%%%%%%%%%%%%%%%%%%%%%%%%%%%%%%%%%%%%%%%
%\begin{figure}[t]
%  \noindent\includegraphics[width=19pc,angle=0]{./figures/figure01.pdf}\\
%  \caption{Enter the caption for your figure here.  Repeat as
%  necessary for each of your figures. Create a figures directory and
%  place all figures in that directory. Figure from Houghton et al. (2001).}\label{f1}
%\end{figure}
%%%%%%%%%%%%%%%%%%%%%%%%%%%%%%%%%%%%%%%%%%%%%%%%%%%%%%%%%%%%%%%%%%%%%
% TABLES
%%%%%%%%%%%%%%%%%%%%%%%%%%%%%%%%%%%%%%%%%%%%%%%%%%%%%%%%%%%%%%%%%%%%%
%
\end{document}